\documentclass[lettersize,journal]{IEEEtran}
\usepackage{amsmath,graphicx,hyperref}
\usepackage{booktabs}
\usepackage[table]{xcolor}
\usepackage[
backend=biber,
style=ieee,
doi=false,
isbn=false,
url=false,
eprint=false,
maxnames=99
]{biblatex}
\usepackage{subcaption}
\usepackage{xcolor}

\begin{document}

\title{An Extensive Analysis of the Singing Voice Conversion Challenge 2025 Evaluation Results}

\author{Lester Phillip Violeta, Xueyao Zhang, Jiatong Shi, Yusuke Yasuda, \\ Wen-Chin Huang, Zhizheng Wu, Tomoki Toda
\thanks{Lester Phillip Violeta and Wen-Chin Huang are with the Graduate School of Informatics, Nagoya University, Japan. \\
Xueyao Zhang and Zhizheng Wu are with the Chinese University of Hong Kong, Shenzhen, China \\
Jiatong Shi is with Carnegie Mellon University, USA \\
Yusuke Yasuda is with the National Institute of Informatics, Japan \\
Tomoki Toda is with Information Technology Center, Nagoya University, Japan \\
}
\thanks{Manuscript received April 19, 2021; revised August 16, 2021.}}

\markboth{Journal of \LaTeX\ Class Files,~Vol.~14, No.~8, August~2021}%
{Shell \MakeLowercase{\textit{et al.}}: A Sample Article Using IEEEtran.cls for IEEE Journals}


\maketitle

\begin{abstract}
We present a thorough analysis of the findings of the latest iteration of the Singing Voice Conversion Challenge, a scientific event aiming to compare and understand different voice conversion systems in a controlled environment. Compared to previous iterations which solely focused on converting the singer identity, this year we also focused on converting the singing style of the singer. To create a controlled environment and thorough evaluations, we developed a new challenge database, introduced two tasks, open-sourced baselines, and conducted large-scale crowd-sourced listening tests and objective evaluations. The challenge was run for two months and in total we evaluated 33 different systems. The results of the large-scale crowd-sourced listening test showed that top systems had comparable singer identity scores to ground truth samples. However, modeling the singing style and consequently achieving high naturalness still remains a challenge in this task, primarily due to the difficulty in modeling dynamic information in breathy, glissando, and vibrato singing styles. Further analyses of the challenge also discuss the limitations of both the traditional similarity test and the dynamic preference test in evaluating singing style similarity. Moreover, calculating Spearman's rank correlation coefficient shows that dependent objective metrics such as chroma-alignment and non-match metrics such as speaker embeddings are the most correlated to subjective scores, but are still not at a level where it could be considered as a true replacement for subjective scores.
\end{abstract}


\section{Introduction}
\label{sec:intro}
Voice conversion (VC) \cite{vc-intro}, the task of converting one kind of speech to another without changing the linguistic contents, has undergone rapid changes recently. This technology has several use cases in the form of entertainment and medical solutions. These use cases range from real-time voice conversion, singing voice conversion, accent conversion, to pathological speech enhancement. Regardless of the wide variety of use cases, the underlying technology of state-of-the-art architectures is almost shared with each other, which either use a parallel sequence-to-sequence \cite{vtn_journal} or recognition-synthesis \cite{s3prl-vc-journal} framework. With this in mind, a benchmarking solution was needed to find the most effective solutions among several methods \cite{vc-intro, vtn_journal, s3prl-vc-journal}.

\begin{figure}[t!]
  \centering
  \includegraphics[width=8.5cm]{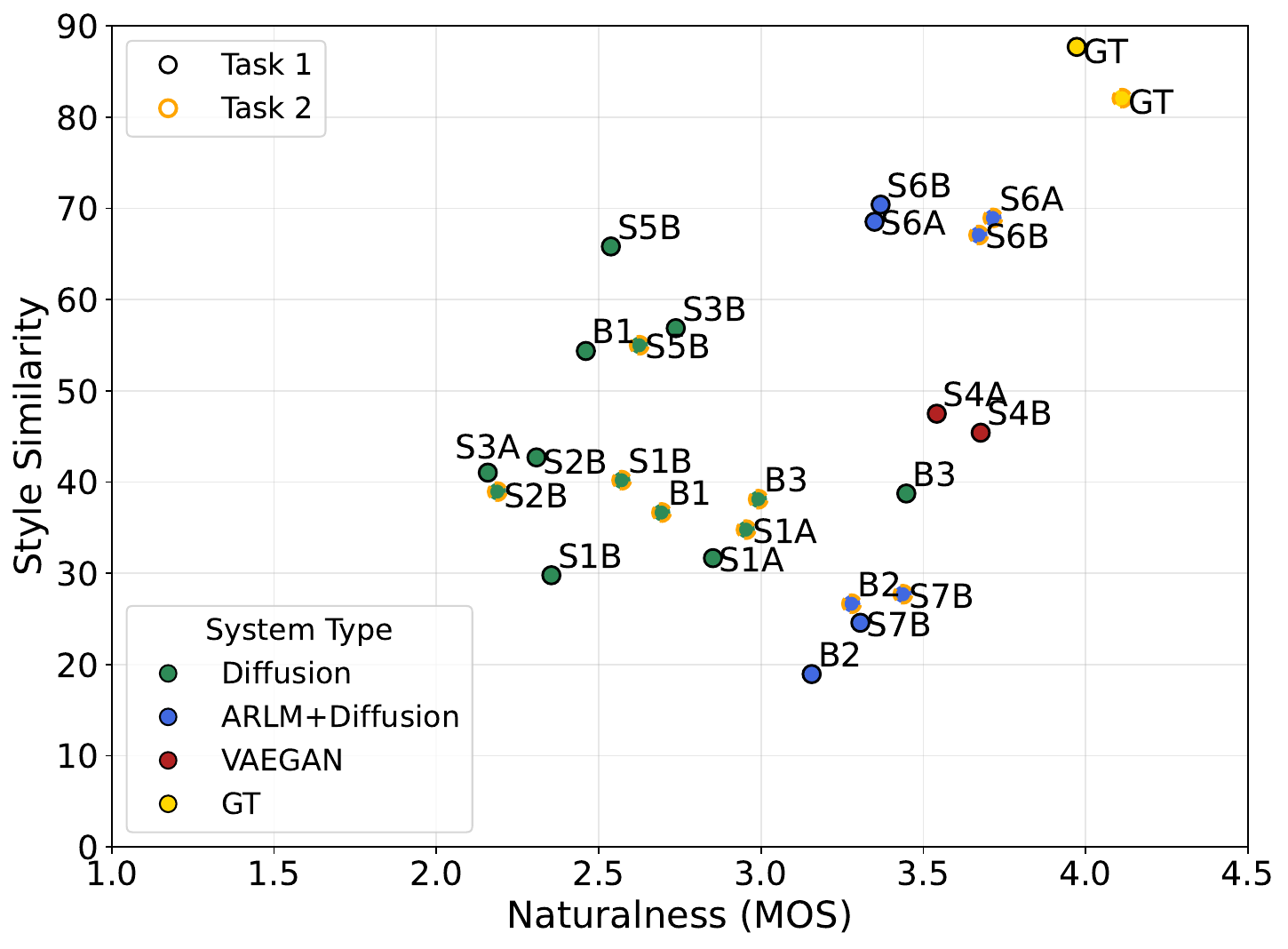}
  \includegraphics[width=8.5cm]{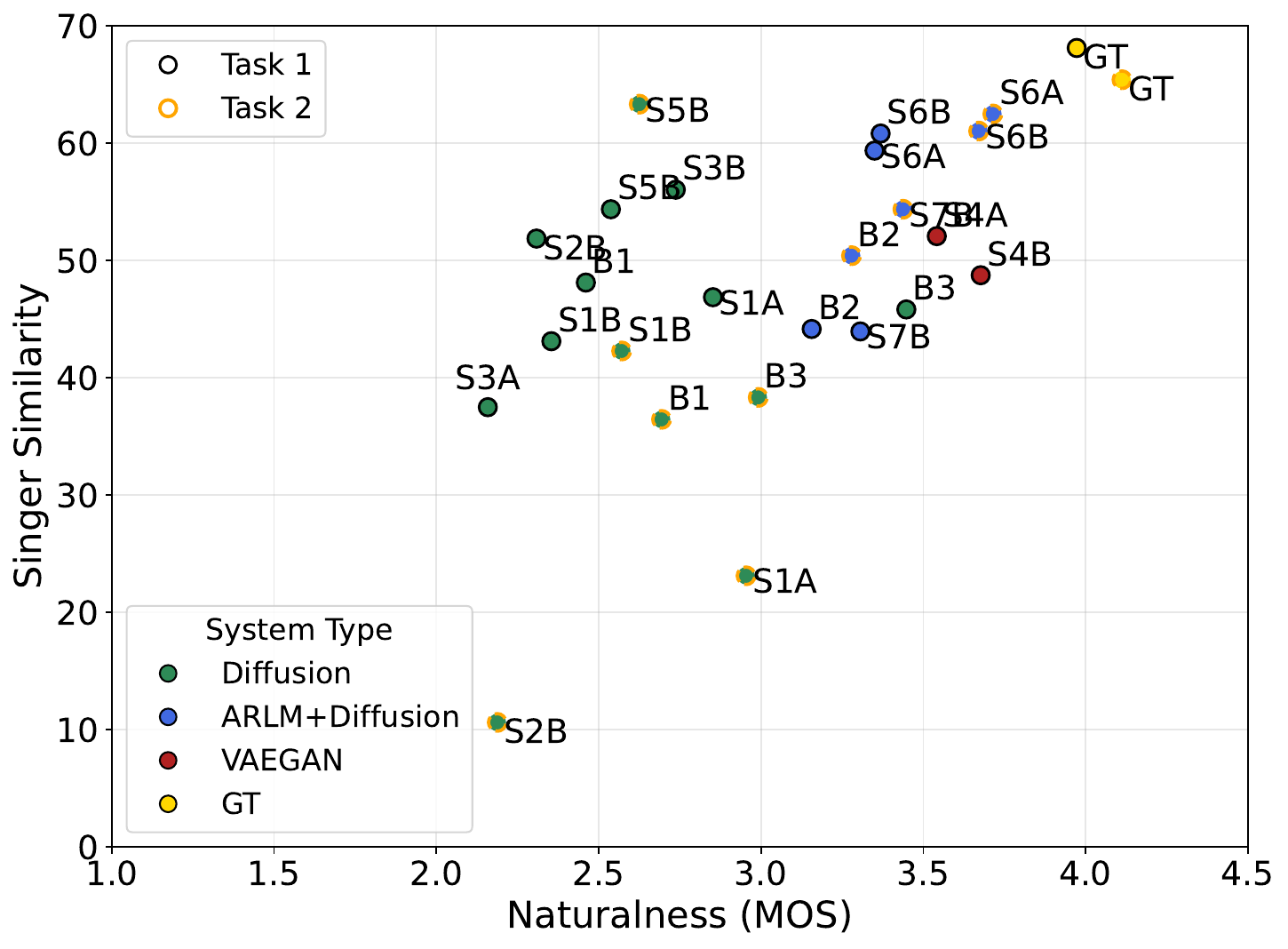}
  \caption{A scatter plot of the results comparing naturalness and style similarity metrics (top) and comparing naturalness and singer identity similarity metrics (bottom) from subjective tests. Higher performing systems are placed on the upper right. Systems labeled with B are teams' best systems, while systems labeled with A are their ablation systems.}
  \label{fig:overall}
\end{figure}

The Voice Conversion Challenge (VCC) series was launched in 2016 \cite{vcc2016} with the goal of comparing several VC approaches in a controlled task setting, with a dedicated database and large-scale evaluation. This was then expanded to more difficult settings such as non-parallel conversion and cross-lingual VC in VCC 2018 \cite{vcc2018} and VCC 2020 \cite{vcc2020}, respectively. Finally, VCC 2020 showed that top systems could already synthesize at a naturalness and similarity score similar to ground-truth recordings. 

With the rapid advancements of speech technology, VCC shifted its focus to singing voices, and Singing Voice Conversion Challenge (SVCC) 2023 \cite{huang2023svcc} was held. Singing voice conversion (SVC) is considered more challenging because of two reasons. First, compared to normal, everyday speech, singing voices have a lot more variation in pitch, energy, expressions, and singing style. Second, while the generated singing voice needs to follow the notes of the song, the singing style can also greatly vary from singer to singer, and thus the level of disentanglement needs to be properly modeled. Two new tasks were introduced, which were in-domain SVC and cross-domain SVC. To evaluate the naturalness and similarity, large-scale listening tests were conducted. For naturalness, the five-point-scale mean opinion score (MOS) test was performed, and for similarity, a four-point-scale speaker-likeness test was performed. The similarity scores were then calculated by the percentage of a system's converted samples that were judged to be the same as the target speakers. Similar to VCC 2020, the large-scale listening tests show human-level naturalness was achieved by the top system, showing the huge advancements in the field of generative speech modeling. In particular, systems employing the VAEGAN \cite{zhou2023t02, ning2023svccwhisper} architecture and DSPGAN-based vocoders \cite{song2023dspgan} performed very well in naturalness. However, unlike VCC 2020, no team was able to obtain a similarity score as high as the target speakers, showing that singing voices may have some intrinsic features that were not considered in the previous challenges.

As singing voices contain more sophisticated and specific features than singer identity, controlling the singing style has more practical applications and needs to be investigated to advance the field of voice synthesis. With this in mind, we launch SVCC 2025. Compared to previous iterations which solely focused on converting the voice identity, this year's iteration also focuses on singing style conversion (SSC). Compared to SVC which is limited to converting the global timbre and static features, SSC is a harder task as both the static and dynamic features of singing voices also need to be controlled. We introduce two new tasks: in-domain SSC and zero-shot SSC and develop a dedicated challenge dataset. Moreover, we provide two open-sourced baseline systems for researchers to use as starter kits to kickstart this new field. Similar to previous years, we conducted large-scale evaluations and summarized the valuable findings in this paper.

We evaluated 33\footnote{The initial paper only analyzed 26 systems, but we included more ablation studies in this paper as explained in Section \ref{sec:rules}.} different systems with submissions from 7 different teams\footnote{Although SVCC 2025 received more than 60 registrations similar to the previous challenges, many registrants did not submit systems due to the novelty of the SSC task and unsatisfactory results at the deadline.}. A quick overview of the results is shown in Fig. \ref{fig:overall}. The results show that, for singer identity, 5 systems had no observed statistically significant difference with the ground truth systems. In terms of singing style and naturalness, no system was able to achieve subjective scores as high as those of the ground truth. The top systems only achieved a naturalness and similarity score of around 3.7 and 70\% respectively, far from the ground truth scores of around 3.9 and 90\%, showing that there is still significant room for improvement. In particular, the lower scores were due to the difficulty in modeling dynamic features in certain singing styles, and consequently resulted in lower naturalness scores. 

Compared to our initial work presented in \cite{svcc2025}, we aim to further extend the discussion to cover the details of the challenge, description of submissions, results and findings of the challenge, and potential future directions. In particular, we discuss the limitations of the traditionally-conducted 4-point similarity test, as the test does not measure the absolute closeness of multiple systems to the target style. In line with this, we discuss the dynamic preference test as an alternative to the traditional XAB similarity test and discuss the strengths and limitations of each evaluation method. In particular, we find that while directly comparing samples can resolve the issues of the traditional XAB similarity test, it may be affected by perceptual factors other than singing style, such as singer
identity. Moreover, we further discuss which objective metrics are most correlated to the subjective scores, and find that dependent and non-match metrics such as chroma-alignment and speaker embeddings are the most correlated. 

Our contributions in this paper are as follows:
\begin{itemize}
  \item We conduct the Singing Voice Conversion Challenge 2025, with a new research task focused on the singing style. A dedicated dataset for the challenge was created, and large-scale subjective tests to evaluate the naturalness, singing style, and singer identity were conducted. Samples of the submitted systems are found in this page \footnote{\url{https://lesterphillip.github.io/svcc2025_demo}}.
  \item We find that Autoregressive Language Modeling (ARLM) + Diffusion-based systems had high performance in the challenge. However, we observe that modeling the singing style and consequently achieving high naturalness still remains a challenge in this task, as the synthesis quality is still not at the same level as the ground truth recordings.
  \item We analyze the subjective scores and discuss both the strengths and weaknesses of the traditional XAB similarity task and an alternative dynamic preference test in evaluating the singing style.
  \item We analyze which objective scores are most correlated with the subjective scores to help researchers analyze their systems. We discover that chroma-alignment-related scores can be correlated to naturalness, and speaker embeddings can also be correlated to the singing style and singer identity. However, there is still work needed to be done before objective tests can become a true replacement for subjective tests, as the correlation scores only reach a maximum score of around 0.8.
\end{itemize}

\begin{table*}[h]
    \normalsize
    \centering
    \begin{tabular}{llll}
        \hline
        Task & Source & Reference & Conversion \\
        \hline
        Task 1 & Singer A, in style 1 & Singer A, in style 2 & Singer A, in style 2 \\
        Task 2 & Singer B, in style 1 & Any singer except B, in style 2 & Singer B, in style 2 \\
        \hline
    \end{tabular}
    \caption{An overview of the tasks conducted in this year's challenge.}
    \label{tab:tasks}
\end{table*}

\section{Related Work}
\subsection{Voice Conversion Challenge Series}
The VCC series started in 2016, where the pilot challenge \cite{vcc2016, vcc2016-analysis} was held as a special session at INTERSPEECH 2016. In the first iteration, 17 participants joined the challenge. To facilitate the challenge, the task focused on parallel VC, and a new database consisting of two source and two target native American English speakers (two females and two males), containing 162 parallel sentences, was constructed for the only task in VCC 2016. Results showed that the best system in VCC 2016 obtained an average naturalness score of 3.0 and a similarity score of 70\%. On the other hand, ground truth recordings had naturalness scores of 4.5 and similarity scores of 90\%. As the difference in naturalness and similarity was far from the ground truth scores, this highlighted the need to continuously improve VC systems.

This was followed by VCC 2018 \cite{vcc2018}, which was held as a special session of the ISCA Speaker Odyssey Workshop 2018, where 32 participants joined the challenge. This focused on two tasks: parallel VC and non-parallel VC. The best system performed well in both parallel and non-parallel tasks and obtained an average of 4.1 in naturalness and about 80\% in similarity. On the other hand, ground truth recordings obtained an average of 4.5 in naturalness and about 90\% in similarity.  Despite significant improvements in architectures and submissions, results showed that there were still statistically significant differences between the ground truth and the best system in terms of both naturalness and speaker similarity.

Finally, the last VCC 2020 \cite{vcc2020} was held in a joint workshop with the Blizzard Challenge, where 30 participants joined the challenge. The challenge focused on two tasks: semiparallel intra-lingual conversion task and cross-lingual conversion task. Results from the listening test showed that for the intra-lingual semi-parallel task, the speaker similarity scores of several systems were as high as the target speakers; however, none of them achieved human-level naturalness. On the other hand, for the cross-lingual task, although the overall naturalness and similarity scores were lower, the best systems had naturalness scores around 4.0 and similarity scores around 90\%, whereas ground truth recordings had naturalness scores of around 4.5 and similarity scores of 95\%.

\subsection{Singing Voice Conversion Challenge 2023}
With the success of submitted systems in the previous challenge, organizers shifted focus to more challenging tasks: in particular, singing voice conversion. SVCC 2023 \cite{huang2023svcc} was held as a special session in IEEE ASRU 2023 where 24 participants joined the challenge. The challenge focused on two tasks: in-domain and cross-domain SVC. Interestingly, listening test results showed that human-level naturalness was achieved by the top system, where both the ground truth recordings and top system achieved around 3.9 naturalness scores. However, no system (top system was at 65\% similarity score) achieved a similarity score with no observed statistical differences compared to the ground truth (80\% similarity score). The challenge highlighted two main points. First, similar to VC, high naturalness was also achievable in SVC, especially with the use of VAEGAN architectures \cite{ning2023svccwhisper, zhou2023t02}. Second, objective evaluation was still difficult to conduct, as most objective metrics were still weakly correlated to the listening test scores.

\section{Timeline, Tasks, Databases, Submission Rules, and Evaluation Methods}
\subsection{Timeline}
The challenge was first announced and promoted on March 10, 2025. Task descriptions and open-sourced baselines were released around the first week of April 2025. Training data was released on April 28, 2025, while the evaluation data was released on June 23, 2025, giving participants around two months to develop their models. Participants were then
asked to submit their converted results on June 30, 2025, and a technical report of their systems by July 21, 2025. 

\subsection{Tasks}
Similarly to previous years, the challenge is composed of two tasks. More details of each task are described below, and an overview of the tasks is described in Table \ref{tab:tasks}.

\noindent{\textbf{Task 1: In-Domain Singing Style Conversion.}} The first task is a straightforward conversion of the same singer into a different singing style. Given the dataset of singer A in different singing styles, the goal during inference is to convert singer A's voice from style 1 into style 2. Since singer A singing in both style 1 and 2 is found inside the training dataset, the goal of this first task is to verify the current capabilities of different systems in conducting this new task.

\noindent{\textbf{Task 2: Zero-Shot Singing Style Conversion.}} The second task involves converting singer B's voice from style 1 into style 2. Different from the first task, although styles 1 and 2 are both still found in the training dataset, singer B's voice data is not found in the training dataset. Thus, the goal of the second task is to evaluate how systems can disentangle the singer identity and singing style features from the singing audio.

\subsection{Databases}
We used a subset of the large-scale and open-sourced dataset called GTSinger \cite{zhang2024gtsinger}, which contains singing data with various parallel song phrases sung in different singing styles and different languages. Other datasets such as SingStyle111 \cite{singstyle111} with various singing style labels were also considered, but organizers opted for the GTSinger dataset due to its large-scale size. During the challenge, participants were asked to handle 7 different styles, namely: breathy, falsetto, mixed voice, pharyngeal, glissando, vibrato, and a control style (only as a source style).

To create the SVCC 2025 dataset, we used the GTSinger dataset and adjusted it as follows. For Task 1's Singer A, we chose the male EN-Tenor-1 singer. We removed 2 songs from singer A and used one of these as the test data. For Task 2's Singer B, we used EN-Alto-2 as the test data. As the goal of Task 2 is to train a zero-shot SSC system, we removed all of Singer B's data from the training set. To ensure that participants would not easily guess the singer in the test data, we also removed EN-Alto-1 from the training data. The dataset is open-sourced\footnote{\url{https://huggingface.co/datasets/lestervioleta/svcc2025}}.

\subsection{Submission rules}
\label{sec:rules}
For this year's challenge, we requested each participant to submit their main system for each task they would join. However, we also allowed participants to voluntarily submit more systems to be evaluated as ablation studies. 
During objective evaluations, all of the submissions (best and ablations) from the participants are evaluated. However, during subjective evaluations, only a maximum of one ablation system is included in the listening tests. Other ablation systems submitted are only evaluated in the objective tests.

For training data, we allowed participants to train on any speech and singing dataset, whether it was open-sourced or privately owned. However, participants were not allowed to use the raw GTSinger dataset as it would result in data leakage. After the challenge, all participants were also required to submit technical reports detailing their architecture and training data. All participants submitted technical reports this year.

\begin{table*}[t]
\centering
\resizebox{\linewidth}{!}{%
\begin{tabular}{|l|l|p{9.5cm}|p{5cm}|}

\hline

\textbf{System} & \textbf{Based on} & \textbf{Fundamental points of best submitted system (and differences from base model)} & \textbf{Datasets} \\ \hline

B1 (Serenade) \cite{violeta2025serenade} & - &
\begin{tabular}[t]{@{}l@{}}
- Disentangled speech features (ContentVec), loudness, and MIDI \\
- Audio infilling training framework with flow-matching  to predict acoustic features 
\end{tabular} &
$\sim$500 hours (singing) \\ \hline
B2 (Vevo1.5) \cite{zhang2025vevo} & - & 
\begin{tabular}[t]{@{}l@{}}
- Two-stage framework with content and content-style tokens \\
- Chromagram-based melody tokenizer
\end{tabular} &
\begin{tabular}[t]{@{}l@{}}
$\sim$100{,}000 hours (speech)\\
$\sim$7{,}000 hours (singing)
\end{tabular} \\ \hline
B3 (NU-SVC) \cite{yamamoto2023svcc} & - &
\begin{tabular}[t]{@{}l@{}}
- Diffusion model developed for SVC task (T13 in SVCC 2023)
\end{tabular} &
$\sim$500 hours (singing) \\ \hline
S1 (Diffusion) & Serenade &
\begin{tabular}[t]{@{}l@{}}
- Decomposed two-stage framework for predicting F0 and \\ mel cepstrum / band aperiodicity features
\end{tabular} &
\begin{tabular}[t]{@{}l@{}}
$\sim$5{,}000 hours (speech)\\
$\sim$500 hours (singing)
\end{tabular} \\ \hline
S2 (Diffusion) & Serenade &
\begin{tabular}[t]{@{}l@{}}
- Use of F0 fluctuation feature
\end{tabular} &
\begin{tabular}[t]{@{}l@{}}
$\sim$75 hours (singing)
\end{tabular} \\ \hline
S3 (Diffusion) & Serenade &
\begin{tabular}[t]{@{}l@{}}
- Content encoder integrating Whisper and HuBERT features\\
- Time-dependent and time-independent style extraction encoders\\
\end{tabular} &
\begin{tabular}[t]{@{}l@{}}
$\sim$75 hours (singing)
\end{tabular} \\ \hline
S4 (VAEGAN) \cite{hu2026controllables4} & SYKI- SVC &
\begin{tabular}[t]{@{}l@{}}
- Whisper feature averaging for style disentanglement and feature fusion\\
- Super-resolution post-processing for style modeling
\end{tabular} &
\begin{tabular}[t]{@{}l@{}}
$\sim$110 hours (singing)
\end{tabular} \\ \hline
S5 (Diffusion) & SeedVC &
\begin{tabular}[t]{@{}l@{}}
- Residual Style Adapter for style encoder \\
- Post-processing with duration alignment and unvoiced processing
\end{tabular} &
\begin{tabular}[t]{@{}l@{}}
$\sim$10{,}000 hours (speech)\\
$\sim$500 hours (singing)
\end{tabular} \\ \hline
S6 (ARLM+Diffusion) \cite{wang2026s6system} & Vevo1.5 &
\begin{tabular}[t]{@{}l@{}}
- Direct Preference Optimization post-training strategy for ARLM \\
- Style-aware cross attention modules
\end{tabular} &
\begin{tabular}[t]{@{}l@{}}
$\sim$100{,}000 hours (speech)\\
$\sim$7{,}500 hours (singing)
\end{tabular} \\ \hline
S7 (ARLM+Diffusion) \cite{zhang2025vevo2} & Vevo1.5 &
\begin{tabular}[t]{@{}l@{}}
- Qwen 2.5 ARLM backbone \\
- Group Relative Policy Optimization post-training strategy for ARLM
\end{tabular} &
\begin{tabular}[t]{@{}l@{}}
$\sim$100{,}000 hours (speech)\\
$\sim$7{,}000 hours (singing)
\end{tabular} \\ \hline
\end{tabular}%
}
\vspace{-6pt}
\caption{Overview of submitted systems, key differences from the respective base models, and training datasets.}
\label{tab:systems}
\end{table*}

\section{Baselines and Submitted Systems}
We describe the baselines and submitted systems used in the large-scale evaluations. An overview of the system name, base architecture, key points, and datasets is shown in Table \ref{tab:systems}. In total, we evaluated 33 different systems, which are composed of 2 ground truth, 6 baselines, and 25 submissions. We categorize the submitted systems into three different categories: VAEGAN, ARLM+Diffusion, and Diffusion.

\subsection{Baseline Systems}
This year, we included three different baseline systems in the evaluation. Two of these were fully open-sourced and released to the participants. All systems were based on the recognition-synthesis framework. Following the definition in \cite{s3prl-vc-journal}, any VC system that separately trains the recognizer and synthesizer can be categorized as this framework.

\textbf{Baseline 1 (B1, Diffusion)} is Serenade \cite{violeta2025serenade}, a diffusion-based model based on an audio infilling architecture. The model uses a masked segment of the target mel-spectrogram and predicts this using a flow-matching model with the complement of the masked target features along with the extracted features. The mel-spectrogram is used as the target acoustic feature and ContentVec \cite{2022qiancontentvec}, loudness, and MIDI information are used as the extracted conditioning features. To disentangle the source singing style, a cyclic training approach where an initial model is used to convert the training data into different styles and trained to reconstruct the original style, is then used. While this model had high singing style similarity scores, the naturalness was quite low due to the source melody not being retained. Thus, another variant was developed to retain the source melody, where a post-processing module using a source-filter-based vocoder was used to resynthesize the converted waveforms using the original F0 patterns, but would result in lower singing style similarity scores. The training and inference code \footnote{\url{https://github.com/lesterphillip/serenade}} was fully open-sourced for participants to use. 

\noindent \textbf{Baseline 2 (B2, ARLM)} is Vevo1.5\footnote{\url{https://github.com/open-mmlab/Amphion/tree/main/models/svc/vevosing}}~\cite{zhang2025vevo} released by the Amphion toolkit~\cite{amphion}, which uses an ARLM+Diffusion architecture. Given the text and melody tokens as input, Vevo1.5 utilizes an auto-regressive transformer to generate the content-style tokens, which are prompted by a singing style reference. To extract melody tokens, Vevo1.5 proposed a chromagram-based melody tokenizer, which is designed to encode only the coarse-grained melody information of the source singing. Then, taking the content-style tokens as input, Vevo1.5 employs a flow-matching transformer to produce mel spectrograms, which are prompted by a timbre reference. The model is pre-trained on large-scale speech and singing voice data to generalize to various styles of singing and conducts zero-shot SSC. 

\noindent \textbf{Baseline 3 (B3, Diffusion)} is based on NU-SVC (SVCC 2023 T13 system) \cite{yamamoto2023svcc}. The model uses a denoising diffusion probabilistic model, along with extracted ContentVec \cite{2022qiancontentvec}, log F0, VUV, and loudness conditioning features to predict the target mel-spectrogram. Then, it uses SiFiGAN \cite{sifigan} to generate the target waveforms. The singer identity was controlled by extracting speaker features from a jointly trained style token encoder. We replaced the original denoising diffusion probabilistic decoder with a flow-matching-based decoder \cite{mehta2024matcha} for more stable training. 

\subsection{Brief description of submitted systems}
\textbf{System 1 (S1, Diffusion)} used Serenade \cite{violeta2025serenade} as the base system. Different from the base system, the submission used a two-stage architecture, where one network first predicts the target dynamic log F0 features and another network predicts static band aperiodicity and mel cepstrum features. To improve Task 2 performance for singer identity similarity, the synthesized results are further processed by NU-SVC \cite{yamamoto2023svcc}. A random subset of 5k hours from the EMILIA \cite{he2024emilia} dataset and various open-sourced singing datasets including the SVCC 2025 training data were used to train the network. Ablation studies where the cyclic training was removed and no NU-SVC post-processing were also submitted. The team submitted for both Tasks 1 and 2, including an ablation study for each task, totaling to 4 submitted systems.

\noindent \textbf{System 2 (S2, Diffusion)} used Serenade \cite{violeta2025serenade} as the base system. However, they used an additional input feature called F0 fluctuation, which is calculated by measuring the offset of the real F0 with the smoothed F0 function (by using Univariate Spline). The system was trained only on the 75 hours of SVCC 2025 training data. The team submitted for both Tasks 1 and 2, totaling to 2 submitted systems.

\noindent \textbf{System 3 (S3, Diffusion)} used Serenade \cite{violeta2025serenade} as the base system. Different from Serenade, a robust content encoder that integrates both HuBERT \cite{hsu2021hubert} and Whisper \cite{radford2023whisper} features based on dual encoding were integrated. Moreover, improvements in the style encoder were added by extracting both time-dependent and time-independent style features. The system was trained only on the 75 hours of SVCC 2025 training data. An ablation study where the robust content encoder was removed was also submitted. The team submitted for Task 1, including an ablation study, totaling to 2 submitted systems.

\noindent \textbf{System 4 (S4, VAEGAN) \cite{hu2026controllables4}} was based on the SYKI-SVC \cite{zhou2025syki} system, and is also based on the SVCC 2023 T02 \cite{zhou2023t02} system, which performed as one of the top systems in the previous challenge. The model is trained in three stages, with each stage focusing on: so-vits-style super-resolution reconstruction, technique disentanglement with Whisper feature \cite{radford2023whisper} averaging, and target singer fine-tuning. The team used around 110 hours of singing data (including the SVCC 2025 training data) to train the model. An ablation study where the Whisper feature averaging strategy was removed (S4A) was also submitted for subjective evaluation, and a system without the use of the super-resolution reconstruction was also submitted for the objective evaluation (S4A1). The team submitted for Task 1, including two ablation studies, totaling to 3 submitted systems.

\noindent \textbf{System 5 (S5, Diffusion)} was based on the open-sourced SeedVC \cite{liu2024seedvc} system, which uses a flow-matching model to generate mel-spectrograms. 
The key difference is the use of the Residual Style Adaptor to capture additional stylistic information, such as articulation and vocal techniques. The team used around 10k hours of speech and 500 hours of singing data including the SVCC 2025 training data to train the model. The team submitted for both Tasks 1 and 2, totaling to 2 submitted systems.

\noindent \textbf{System 6 (S6, ARLM+Diffusion) \cite{wang2026s6system}} was based on Vevo1.5 \cite{zhang2025vevo2} and was improved by taking the provided pretrained weights from Baseline 2, and fine-tuning the ARLM part using Direct Preference Optimization (DPO) \cite{rafailov2023dpo} post-training, by first fine-tuning with LoRA \cite{hu2022lora}, and introducing FiLM and style cross-attention modules into the ARLM. The researchers also used 500 hours of carefully filtered web-scraped data and the SVCC 2025 training data to fine-tune both the ARLM and the flow-matching model. An ablation study where the ARLM was instead fine-tuned without the use of DPO (S6A) was also submitted for subjective evaluation. Other ablation studies for objective evaluation includes: full parameter fine-tuning (S6A1), LoRA fine-tuning only (S6A2), and LoRA fine-tuning with only FiLM embeddings (S6A3). The team submitted for both Tasks 1 and 2, including four ablation study for each task, totaling to 10 submitted systems.

\noindent \textbf{System 7 (S7, ARLM+Diffusion) \cite{zhang2025vevo2}} was based on Vevo1.5 \cite{zhang2025vevo2} but introduced the use of the Qwen 2.5-0.5B \cite{Yang2024Qwen25report} as the ARLM, and was further trained with the Group Relative Policy Optimization \cite{shao2024deepseekmath} post-training strategy. The researchers used the same datasets in Vevo1.5 and the SVCC 2025 training data to train the model. The team submitted for both Tasks 1 and 2, totaling to 2 submitted systems.

\section{Subjective evaluations}
For subjective evaluation, we conduct a large-scale listening test and evaluate naturalness, singing style similarity, and singer identity similarity. All of the audio is in 24 kHz and normalized using sv56\footnote{\url{github.com/openitu/STL/tree/dev/src/sv56}}.

\subsection{Evaluation protocol}
We first use the conventional evaluation methods that were used in the previous challenges \cite{huang2023svcc}. To evaluate naturalness, a mean opinion score (MOS) test was conducted, where listeners were asked to evaluate on a five-point scale. For style similarity, an XAB test was conducted, where listeners were presented with three parallel samples: X (the converted or ground truth utterance), A (the ground truth source sample, pitch shifted), and B (the ground truth target sample, pitch shifted)\footnote{When X is the ground truth, X and B would become the same, thus we applied pitch shifting to avoid bias during the listening test as was conducted in \cite{violeta2025serenade}.}. For singer identity similarity, three natural singing voice samples from the target singer and a converted singing sample were presented, and listeners were asked to judge whether the converted samples were produced by the target singer on a four-point scale. Each system received 480 ratings.

\vspace{-6pt}
\begin{figure}[t!]
  \centering

  \begin{subfigure}[t]{0.95\linewidth}
    \centering
    \includegraphics[width=\linewidth]{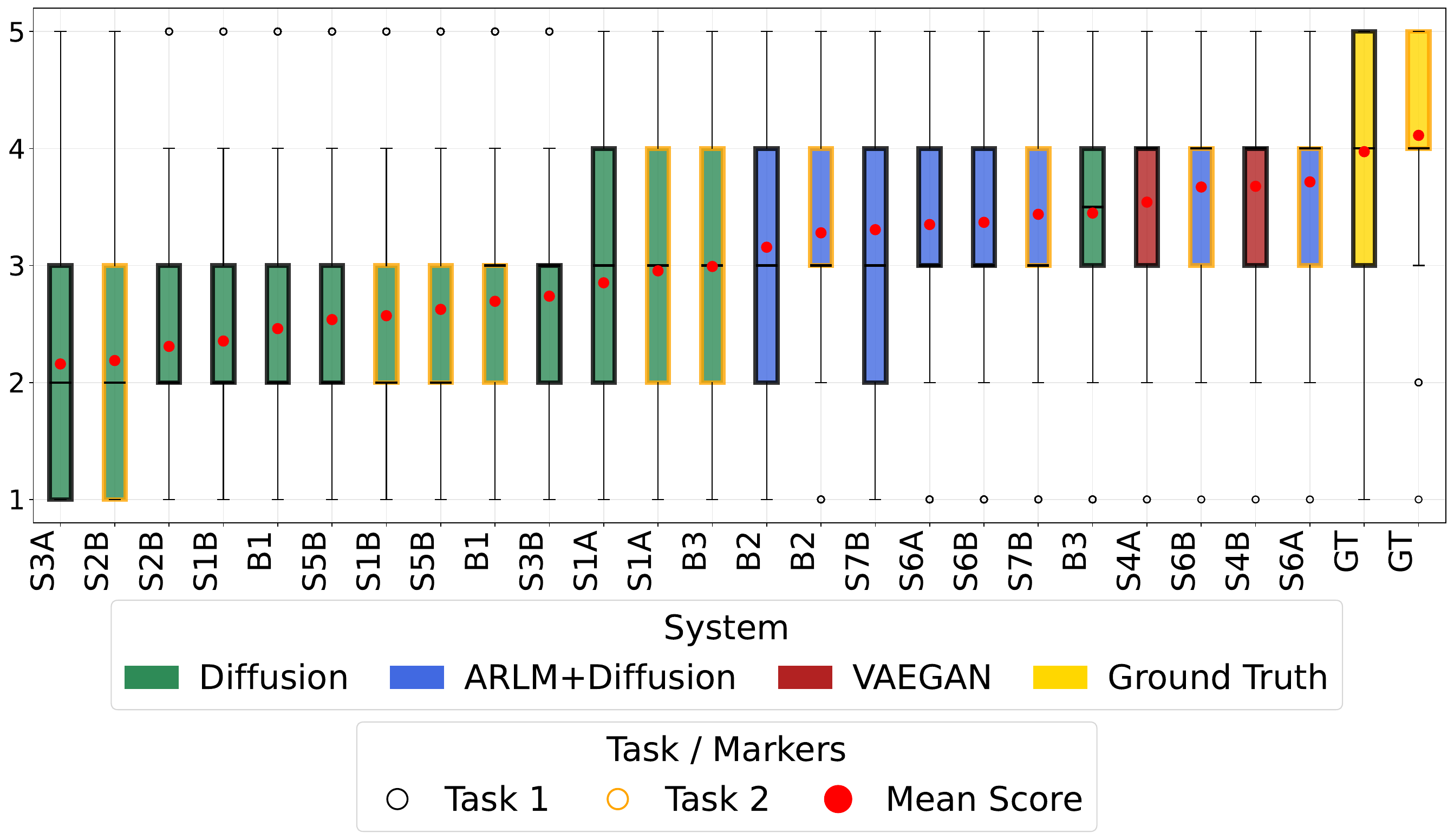}
    \caption{Naturalness MOS}
    \label{fig:results:naturalness}
  \end{subfigure}

  \vspace{0.6em}

  \begin{subfigure}[t]{0.95\linewidth}
    \centering
    \includegraphics[width=\linewidth]{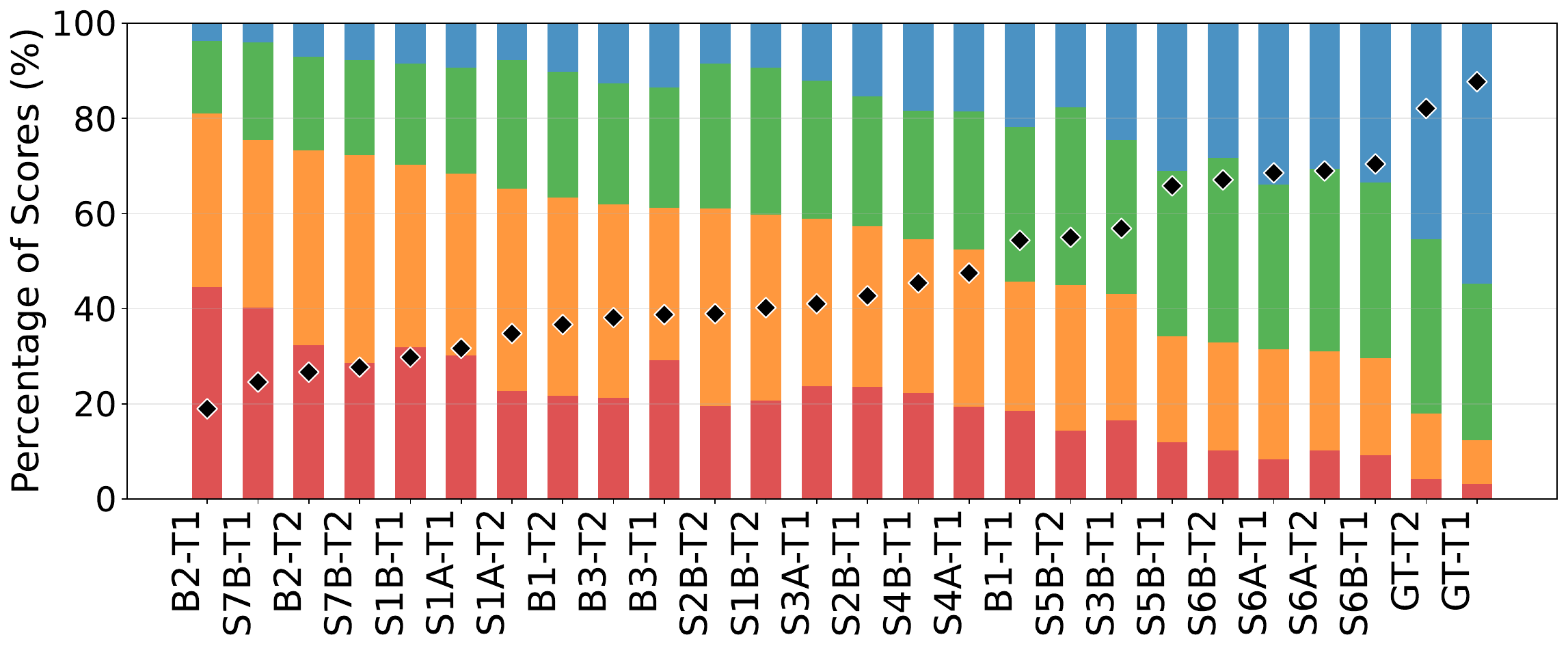}
    \caption{Singing style similarity}
    \label{fig:results:style}
  \end{subfigure}

  \vspace{0.6em}

  \begin{subfigure}[t]{0.95\linewidth}
    \centering
    \includegraphics[width=\linewidth]{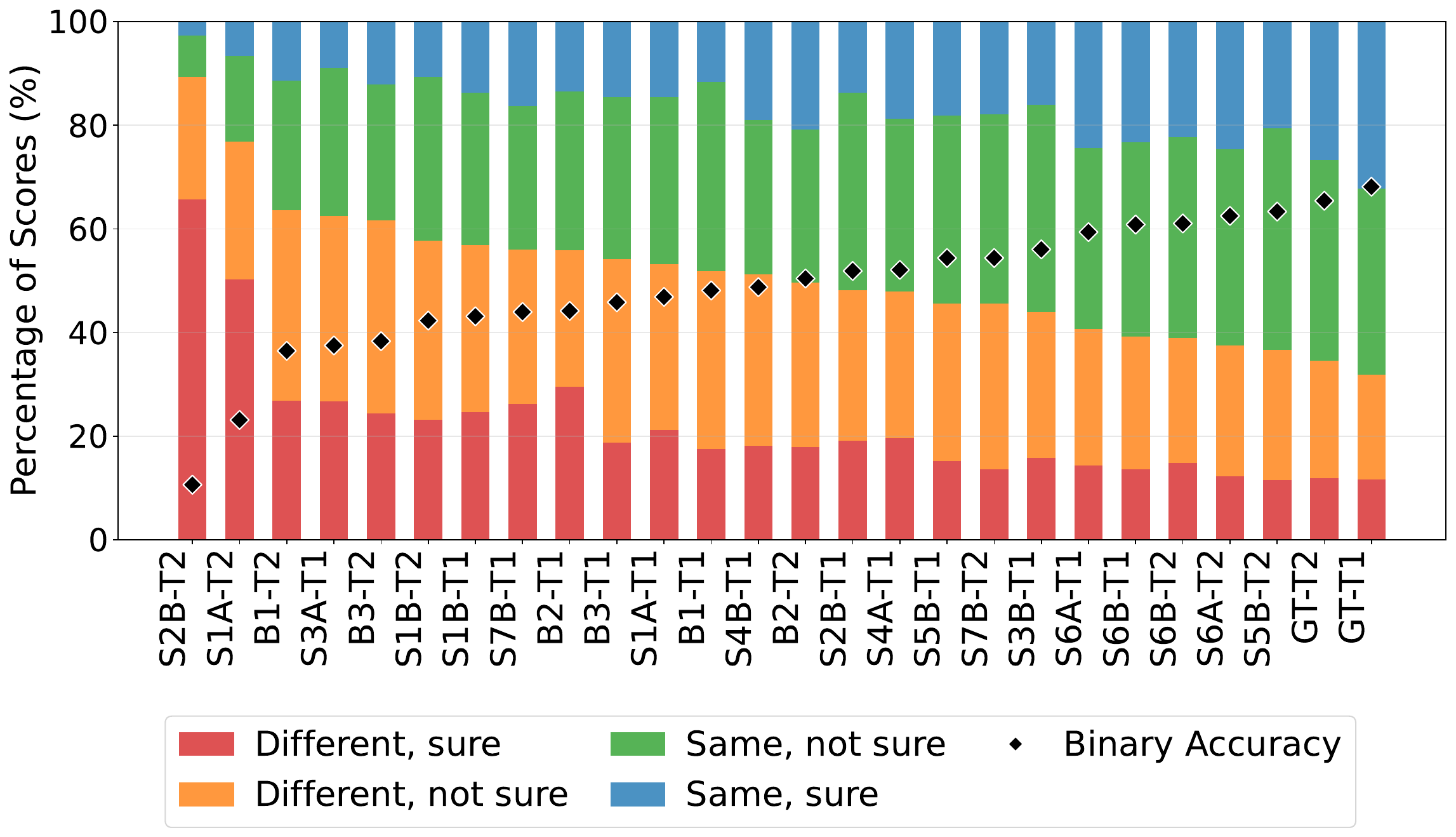}
    \caption{Singer identity similarity}
    \label{fig:results:identity}
  \end{subfigure}

  \caption{Boxplots of the naturalness, singing style similarity, and singer identity similarity. Systems are arranged from right (highest) to left (lowest) based on average scores.}
  \label{fig:results}
\end{figure}

\subsection{Evaluation results}
\subsubsection{Naturalness tests}
Fig.~\ref{fig:results} shows the subjective evaluation results.
First, for naturalness, no system achieved human-level naturalness, unlike in the previous challenges. We attribute this to the sophisticated requirements of also converting the singing style, which complicated some architectures. Top performing architectures in SVCC 2023 such as S4 (VAEGAN-based) and even decent performing architectures B3 (diffusion-based) still performed well in this challenge. However, we observe that new architectures such as ARLM+Diffusion also consistently performed well. 

\subsubsection{Singing style similarity tests}
In terms of singing style similarity, there was a larger gap between the ground truth and the top systems, showing the amount of work that needs to be done in this field. The top system (S6) was achieved by fine-tuning the ARLM part of Vevo1.5 either by LoRA or DPO, and pure diffusion-based models using SeedVC and Serenade were behind this system. We investigated which singing styles were the hardest to model by the different systems. We found that the difficult singing styles to model were breathy, glissando, and vibrato, achieving only 37.3\%, 42.6\%, and 43.9\% style accuracy scores, respectively. On the other hand, pharyngeal, mixed voice, and falsetto singing styles were the easiest to model, with 44.8\%, 48.3\%, and 48.8\% style accuracy scores, respectively. We suspect that breathy, glissando, and vibrato voices have strong dynamic components (time-varying noise in breathy, rapid F0 modulation in vibrato, continuous F0 drift in glissando), which have not been as focused on in previous SVC works, while pharyngeal, mixed voice, and falsetto voices mainly contain specific static components.

\subsubsection{Singer identity similarity tests}
In terms of singer identity similarity, 5 systems have no observed statistically significant difference with the ground truth scores, showing advancements compared to the previous SVCC 2023. In particular, S5B-T2 and all systems from S6 achieved the highest scores. We also attribute these improvements to the listening test setup, where instead of simply showing one ground truth reference sample of the singer (like in the previous challenges), we now show three reference samples of the singer, such that listeners have a more general idea of the singer's voice. However, we acknowledge that singer identity is still difficult to evaluate, as even ground truth samples only achieved a similarity score of around 70\%.

\subsection{Preference test protocol}

In addition to the MOS test, we experiment with preference-based scoring evaluation methods to further evaluate the naturalness and singing style similarity. As the quality of synthetic speech gets closer to human-level recordings, it becomes more difficult for listeners to accurately evaluate speech in a five-point scale as two different systems may contain different artifacts but also have the same overall sound quality. Compared to the MOS test which is a direct scoring method, preference-based methods can compare two audio samples from different systems and instead just allow listeners to choose the better system. This helps remove the bias in direct scoring evaluations when scoring other systems and simplifies the evaluation method. To evaluate the best systems in the preference test, we calculate the win rate of one system compared to another system by calculating the win counts over the number of comparisons. Different from the MOS test, we separate Tasks 1 and 2 into separate tests to make the singer identity uniform throughout comparisons, as the difference in the singer identity would introduce more variables and make it more difficult to evaluate the better system.

A downside to preference-based testing is that the amount of evaluations needed significantly increases as each system needs to be compared with each other, thus requiring $n$ systems in a single test to have $O(n^2)$ evaluations. Moreover, in large-scale evaluation settings such as SVCC where there are multiple systems to be evaluated, more listeners need to be recruited and longer evaluation tests need to be conducted, which becomes increasingly expensive. To resolve this, we use an online learning method called MERGE-RANK algorithm (MRA) \cite{yasuda_preftest} to optimize the combination of pairs. MRA compares the minimum number of pairs required to determine the total order of evaluation targets, while also ensuring evaluation errors are below threshold. Initial tests using MRA to evaluate the results of the previous SVCC 2023 challenge showed that MRA successfully optimizes the combination of pairs and reduced the number of pair combinations from 351 to 83 without compromising evaluation accuracies and wasting budget allocations. Due to resource limitations, we only conduct additional preference tests for the naturalness and singing style evaluations. The tolerance bias and confidence level were set to $\epsilon = 0.0961$ and $\delta = 0.05$, respectively. Under this setting, the maximum evaluation budget required to determine a winner for each pair was $m = 200$.

\subsection{Preference test results and analyses}
We compare the results of the MOS test with the preference tests. First, we observe the difference of how systems were ranked in the direct scoring method and the preference test.

\begin{figure}[t!]
  \centering
  \includegraphics[width=10cm]{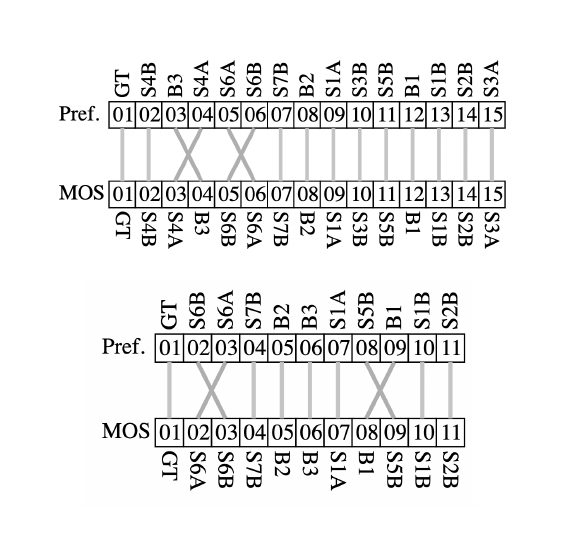}
  \caption{A comparison of the rankings of each system in the MOS test and the preference test when evaluating naturalness. Task 1 is shown at the top and Task 2 is shown at the bottom.}
  \label{fig:mos_pref}
\end{figure}

\begin{figure}[t!]
  \centering
  \includegraphics[width=10cm]{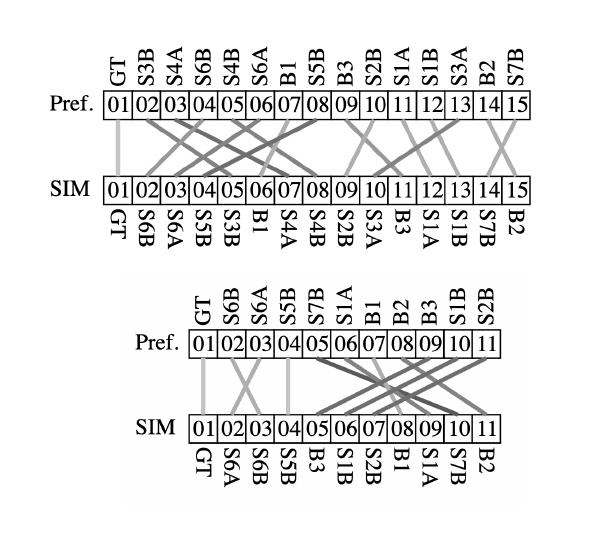}
  \caption{A comparison of the rankings of each system in the similarity test and the preference test when evaluating style similarity. Task 1 is shown at the top and Task 2 is shown at the bottom.}
  \label{fig:sim_pref}
\end{figure}

We acquire the final rankings of the system based on their win rate \cite{yasuda_preftest}. For naturalness, as seen in Fig. \ref{fig:mos_pref}, only four systems in both Tasks 1 and 2 switched positions when evaluating the naturalness, showing that the direct scoring and preference tests are highly correlated. Moreover, due to the high variation of audio quality in the submitted systems, evaluating naturalness became quite easy for the listeners. Nevertheless, the use of the preference test for evaluating the naturalness of multiple systems would still hold value in the future, especially once the synthesis quality becomes higher and the gap between each system becomes smaller. On the other hand, for style similarity, Fig. \ref{fig:sim_pref} shows a different story, as the rankings completely change. 

\begin{figure}[t!]
  \centering
  \includegraphics[width=8.5cm]{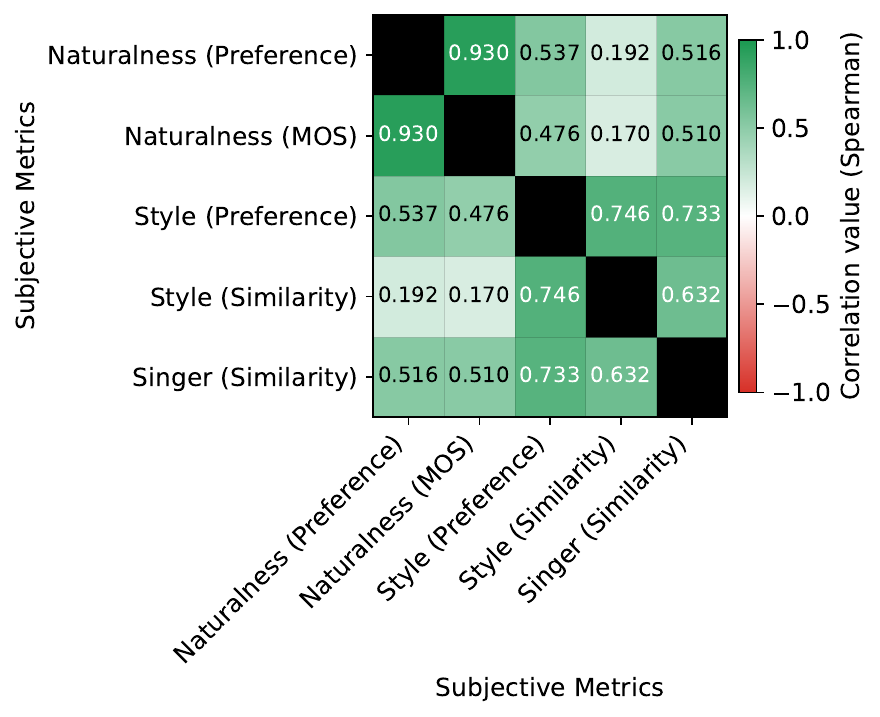}
  \caption{A visualization of the Spearman's rank correlation coefficient between different subjective metrics.}
  \label{fig:subj_vs_subj}
\end{figure}

This change of the final rankings is also observed when calculating the Spearman's rank correlation coefficient (SRCC) of each subjective metric as visualized in Fig. \ref{fig:subj_vs_subj}. The naturalness scores from the preference and MOS test are highly correlated at around 0.930, while the style scores are slightly less correlated at only 0.746. Even more interestingly, the singing style scores from the preference test are more correlated to naturalness scores from both the preference and MOS test at 0.537 and 0.476 respectively, compared to the singing style scores from the similarity test at only 0.192 and 0.170. Moreover, the singer identity scores are more correlated to the style when using the preference test at 0.733, rather than the similarity test at only 0.632. We attribute this difference in results to the difficulty of evaluating the singing style compared to naturalness, as there is an increased amount of subjectivity in music. Moreover, this difference can also be attributed as to how the evaluations are conducted. 

\subsection{Discussion of limitations of traditional similarity test}
With the results from both the direct scoring and preference tests, an important question that remains is how the subjective evaluation results should be interpreted. To answer this question, we discuss the limitations of the traditional similarity (XAB) test for evaluating singing style similarity and compare them with those of the dynamic preference test. In the traditional XAB similarity test, listeners are given a converted sample X together with two references: the source singing style A and the target singing style B. They are then asked to judge whether X is closer to A or B. Therefore, the task can be regarded as a binary preference judgment between the two references, with additional confidence levels represented by the four response choices. However, this setting can still be problematic when the goal is to compare multiple systems in terms of their absolute closeness to the target style.

The main limitation of the traditional similarity is that the XAB score reflects the relative closeness of a converted sample to the target style compared with the source style, rather than directly comparing the distances of different systems from the target style. For example, suppose that two converted samples X1 and X2 are obtained from two different systems. Even if X1 is closer to the target style B than X2 in terms of absolute distance, X2 may receive a better similarity score if it is much farther from the source style A and is therefore judged as relatively closer to B than to A. In this case, the XAB test may rank X2 higher than X1, although X2 is not actually closer to the target reference.

\begin{figure}[t!]
  \centering
  \includegraphics[width=9cm]{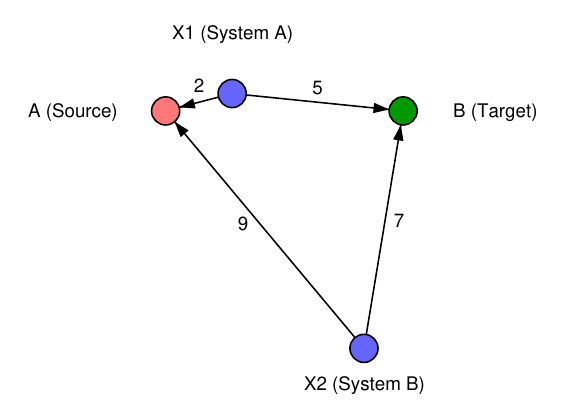}
  \caption{A visualization of a potential limitation of the traditional XAB similarity test. In this setting, listeners judge whether the converted sample is closer to the source reference or the target reference, with additional confidence levels represented by the four response choices. The resulting score reflects the relative closeness of each sample to the two references, but it does not directly compare the absolute distances of different systems from the target reference.}
  \label{fig:pref_adv}
\end{figure}

This limitation is illustrated in Fig. \ref{fig:pref_adv}. In this example, X1 is closer to the target style B than X2 in terms of absolute distance, because the distance between X1 and B is smaller than the distance between X2 and B. However, X2 is farther from the source style A and is still closer to B than to A. Therefore, in the traditional XAB similarity test, X2 may receive a higher score than X1 because it is judged as relatively closer to the target than to the source, even though X2 is not the closest sample to the target style. This illustrates that the XAB score does not directly compare the absolute target-style similarity of different systems. One possible way to address this issue is to directly compare the distances of different systems from the target reference. For example, even within an XAB-style framework, the target style could be used as the reference X, while A and B could be converted samples generated by two different systems. The listener would then directly judge which converted sample is closer to the target style. This setting would better reflect the intended comparison between systems. However, when the number of submitted systems increases, the number of required pairwise comparisons grows rapidly, making such an exhaustive evaluation impractical for a large-scale challenge.

In such a case, the dynamic preference test provides a more practical alternative to this exhaustive pairwise comparison. Instead of requiring all possible system pairs to be evaluated, it adaptively selects pairs of samples and estimates the relative ranking of systems from the collected preferences. In this way, it can more directly compare systems in terms of perceived closeness to the target style while keeping the evaluation size manageable. 

However, the current setting of the dynamic preference test also has a potential limitation. As listeners are asked to judge which converted sample is closer to the target singing style, their judgments may be influenced not only by singing style similarity but also by singer identity similarity. This is especially problematic because both singing style similarity and singer identity similarity are perceptually subtle and may be difficult for listeners to clearly separate. Therefore, a system that better preserves or converts singer identity could receive a higher preference score even if its singing style similarity is not necessarily better.

On the other hand, the traditional similarity test may partially alleviate this problem by presenting both the source and target references. These two references can help listeners infer which perceptual attributes should be emphasized in the judgment, and may make it easier to distinguish singing style similarity from other factors such as singer identity. Thus, although the XAB test has limitations in comparing the absolute closeness of different systems to the target style, it may provide clearer contextual cues for listeners.

Based on these observations, it is difficult to conclude at this stage that one evaluation method is strictly better than the other. The traditional XAB similarity test may be limited when comparing the absolute closeness of multiple systems to the target style, while the dynamic preference test may be affected by perceptual factors other than singing style, such as singer identity. These results suggest that evaluation methodology for singing style similarity remains an open issue.

Future work should investigate evaluation designs that can more clearly separate singing style similarity from singer identity similarity while still allowing scalable comparison among many systems. One possible direction for future work is to conduct an additional dynamic preference test using the source style as the reference and analyze the relationship between the results obtained with the source-style and target-style references. Such an analysis may help clarify whether listeners' judgments are primarily driven by singing style similarity, singer identity similarity, or other perceptual factors. However, this requires further experimental design and is left for future investigation.

\section{Objective evaluations}
Developing systems which can evaluate the naturalness \cite{huang2024voicemos} and singer identity \cite{torres2023_singeremb} similar to humans is an active research area but there have been no previous works yet to evaluate the accuracy of the singing style. However, although subjective tests are considered to be the gold standard in evaluating speech and singing audio, these can be expensive to conduct during initial tests. For example, due to cost and listener fatigue, subjective tests were strictly limited to a maximum of one ablation system per team. Objective tests, on the other hand, enabled a comprehensive and exhaustive assessment of all 26 systems (including all omitted ablation systems) across the entire dataset. Also, while human listening tests tend to be holistic, objective metrics provide high resolution by separating and quantifying specific acoustic features (e.g., pitch via CENS/CQT, timbre via speaker embeddings). This allows for a detailed breakdown of why a system was perceived as good. Thus, although objective tests may not be seen as accurate as subjective tests in evaluating synthetic singing voices, the aforementioned reasons make objective tests be an area that also needs to be focused on.

\subsection{Objective test protocol}

For objective tests, we use VERSA \cite{shi2025versa} for a comprehensive evaluation of speech signals. We use all metrics available in the toolkit that can be used to evaluate the different tasks. Specifically, metrics\footnote{We use the categories as defined in the VERSA \cite{shi2025versa} paper.} such as independent metrics (Sheet SSQA \cite{huang25g_sheetssqa}, SingMOS \cite{tang2025singmos}, AudioBox Aesthetics \cite{tjandra2025aesthetics}), dependent metrics (chroma-related alignment using STFT, CENS, and CQT), non-match metrics (speaker embeddings \cite{jung24c_speakeremb}, singer embeddings \cite{torres2023_singeremb}) and distributional metrics (FAD \cite{gui2024fad}, Audio Density and Coverage \cite{naeem2020audiodistribution}) are used to ensure variation. Finally, we elaborate a correlation analysis to find which ones are most correlated to the subjective scores. As not all ablation systems were used in the subjective evaluations, we provide a final ranking of all systems for each subjective test based on the most correlated evaluation metric.

\subsection{Can objective metrics predict subjective ratings?}
\begin{figure}[t!]
  \centering
  \includegraphics[width=7.5    cm]{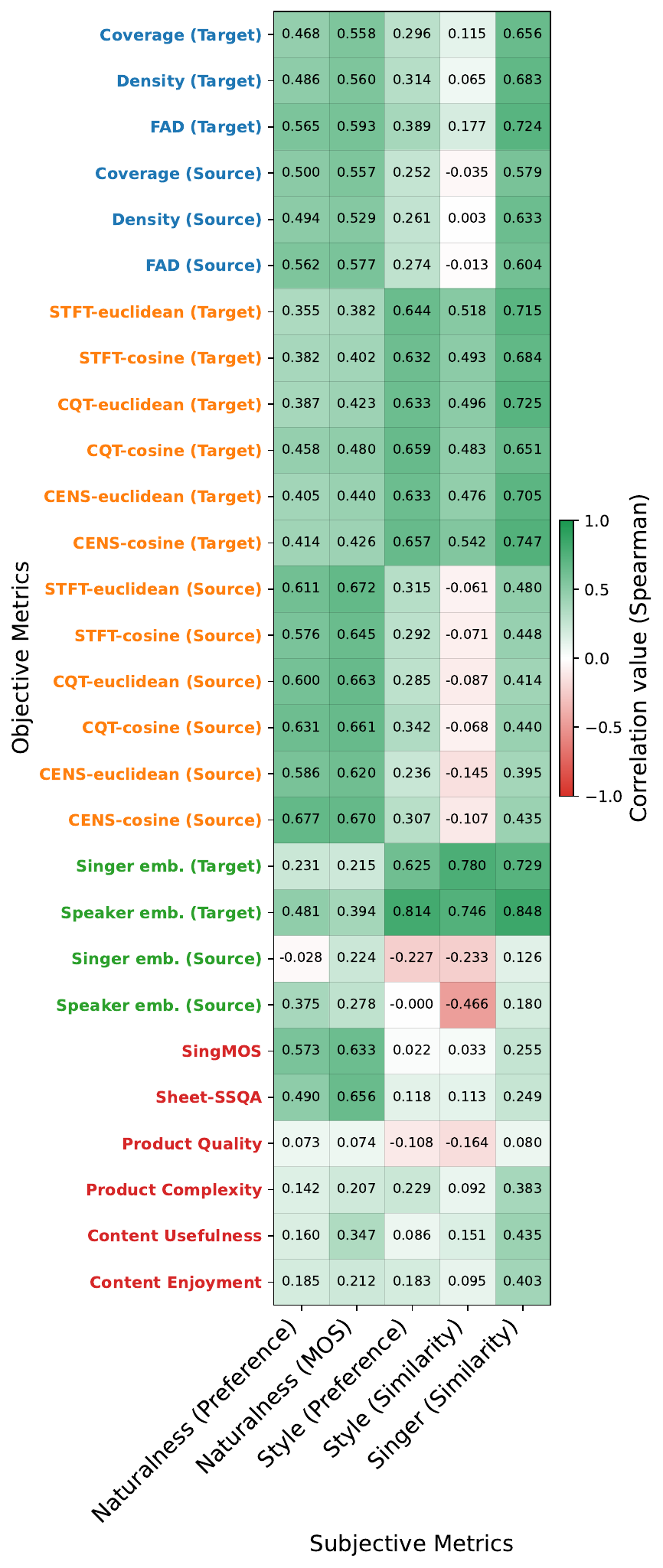}
  \caption{A visualization of the correlations of objective tests with subjective tests. Each objective metric is colored based on its evaluation type. Blue text indicates distributed metrics, orange text indicates dependent metrics, green text indicates non-match metrics, and red text indicates independent metrics.}
  \label{fig:subj_obj}
\end{figure}

We first focus on which metrics are the most correlated to the subjective tests. To achieve this, we calculated each metric's SRCC with the subjective scores. An overview of all the results can be found in Fig. \ref{fig:subj_obj}.

For naturalness, we initially found that neural-based MOS system SingMOS \cite{tang2024singmos} had the highest correlation with subjective scores in both the naturalness preference and MOS tests. However, we also found that chroma-alignment scores with CENS source features were also correlated to the naturalness, and even more correlated to the independent metrics. This also makes sense, primarily because naturalness in singing audio is also correlated to how the converted utterance retains the source melody of the singing voice. If the singing voice becomes out-of-tune, the naturalness also decreases, even in cases where there are minimal artifacts from computer processing.

On the other hand, for both singer identity and singing style similarity scores, using speaker embeddings \cite{jung24c_speakeremb} had the highest correlation with the subjective scores. However, we also see a similar case to naturalness, where the chroma-alignment-related metrics with CQT and CENS target features are correlated to the subjective scores. This is also aligned with expectations, as the ground truth target contains both the singer identity and singing style that needs to be modeled.

These findings further show the importance of conducting academic challenges with controlled datasets to properly benchmark different systems. For example, independent metrics and non-match metrics are useful in real-world cases when the ground-truth utterance is not available. However, in academic challenges such as SVCC where the ground-truth utterance is available, dependent metrics would have a higher correlation to subjective tests, allowing researchers to easily benchmark their systems. However, in both cases, the correlation scores between the subjective and objective tests are quite low and objective scores still have a long way to go before they can be used as a true alternative. Although improving objective evaluations is an active research topic and several research advancements have been made, more work still needs to be done to improve their correlation with subjective metrics. Nevertheless, using these objective scores can be a cheap and quick way for researchers to benchmark their systems before conducting expensive subjective tests.

\subsection{Ranking of all submitted systems based on objective scores}

\begin{table*}[t]
\centering
\caption{System ranking (including all ablation systems) based on objective metrics that correlate the most to naturalness, singing style, and singer identity subjective scores. Lower average rank indicates better overall performance. System names are highlighted by model category. Blue indicates ARLM+Diffusion, red indicates VAEGAN, and green indicates pure diffusion.}
\label{tab:objective_rank_table}
\setlength{\tabcolsep}{3.5pt}
\begin{tabular}{lrrrrrrr}
\toprule
System & CENS cos. (Src.) & CQT cos. (Src.) & Spk. emb. (Tgt.) & Sing. emb. (Tgt.) & CENS cos. (Tgt.) & CQT cos. (Tgt.) & Average rank \\
\midrule
\cellcolor{blue!15}S6A1-T2 & 8 & 3 & 2 & 5 & 4 & 4 & 4.33 \\
\cellcolor{blue!15}S6A-T2  & 11 & 8 & 2 & 5 & 2 & 2 & 5.00 \\
\cellcolor{blue!15}S6A2-T2 & 10 & 7 & 2 & 5 & 3 & 3 & 5.00 \\
\cellcolor{blue!15}S6B-T2  & 12 & 9 & 2 & 5 & 1 & 1 & 5.00 \\
\cellcolor{blue!15}S6A3-T2 & 9 & 5 & 2 & 5 & 5 & 5 & 5.17 \\
\cellcolor{blue!15}B2-T2   & 6 & 4 & 21 & 16 & 6 & 6 & 9.83 \\
\cellcolor{red!15}S4A-T1   & 1 & 1 & 10 & 22 & 15 & 12 & 10.17 \\
\cellcolor{blue!15}S6A-T1  & 16 & 19 & 8 & 1 & 9 & 10 & 10.50 \\
\cellcolor{blue!15}S6A2-T1 & 15 & 15 & 10 & 1 & 11 & 11 & 10.50 \\
\cellcolor{blue!15}S6B-T1  & 18 & 20 & 8 & 1 & 8 & 9 & 10.67 \\
\cellcolor{red!15}S4B-T1   & 2 & 2 & 13 & 19 & 17 & 15 & 11.33 \\
\cellcolor{green!15}S3B-T1 & 21 & 21 & 1 & 13 & 7 & 8 & 11.83 \\
\cellcolor{red!15}S4A1-T1  & 3 & 6 & 13 & 16 & 19 & 20 & 12.83 \\
\cellcolor{blue!15}S6A1-T1 & 13 & 13 & 15 & 5 & 18 & 16 & 13.33 \\
\cellcolor{blue!15}S6A3-T1 & 14 & 14 & 15 & 5 & 20 & 18 & 14.33 \\
\cellcolor{green!15}S5B-T2 & 22 & 22 & 10 & 1 & 14 & 19 & 14.67 \\
\cellcolor{green!15}S5B-T1 & 27 & 26 & 7 & 5 & 12 & 17 & 15.67 \\
\cellcolor{green!15}S2B-T1 & 23 & 23 & 20 & 16 & 12 & 13 & 17.83 \\
\cellcolor{blue!15}S7B-T1  & 4 & 10 & 24 & 27 & 21 & 22 & 18.00 \\
\cellcolor{blue!15}S7B-T2  & 4 & 10 & 24 & 27 & 21 & 22 & 18.00 \\
\cellcolor{green!15}B1-T1  & 25 & 24 & 15 & 19 & 16 & 14 & 18.83 \\
\cellcolor{green!15}S1A-T2 & 20 & 17 & 30 & 29 & 10 & 7 & 18.83 \\
\cellcolor{blue!15}B2-T1   & 7 & 12 & 24 & 30 & 23 & 21 & 19.50 \\
\cellcolor{green!15}S1A-T1 & 17 & 16 & 18 & 19 & 26 & 26 & 20.33 \\
\cellcolor{green!15}S1B-T1 & 19 & 18 & 21 & 24 & 27 & 27 & 22.67 \\
\cellcolor{green!15}B3-T1  & 26 & 27 & 18 & 22 & 24 & 24 & 23.50 \\
\cellcolor{green!15}S3A-T1 & 23 & 25 & 27 & 26 & 25 & 25 & 25.17 \\
\cellcolor{green!15}S1B-T2 & 28 & 28 & 21 & 24 & 27 & 27 & 25.83 \\
\cellcolor{green!15}B3-T2  & 30 & 30 & 28 & 14 & 30 & 29 & 26.83 \\
\cellcolor{green!15}B1-T2  & 31 & 31 & 29 & 14 & 31 & 31 & 27.83 \\
\cellcolor{green!15}S2B-T2 & 29 & 29 & 31 & 30 & 29 & 29 & 29.50 \\
\bottomrule
\end{tabular}
\end{table*}

Finally, we also analyze all the submitted systems and create an overall ranking based on the most correlated objective scores. We use the results from the preference test for naturalness and singing style, and use the results from similarity test for the singer identity to find the top two most correlated objective metrics. For naturalness, we use the CENS-cosine and the CQT-cosine with respect to the source utterance. For singing style (preference), we use the speaker embedding and CQT-cosine with respect to the target utterance, and singing style (similarity), we use the speaker embedding and singer embedding. Finally, for singer identity, we use the speaker embeddings and the CENS-cosine with respect to the target utterance. We take the ordinal ranking of each system based on their objective scores and take their average ranking. Moreover, since the speaker embeddings were the most correlated for both singing style and singer identity, we double its weight when calculating the average rank. However, it must be emphasized that these are not the final rankings, as the SRCC between objective and subjective scores do not show a perfect correlation and only reach a maximum of 0.8 SRCC. Nevertheless, it would be useful to discover where the ablation systems scored to further improve future systems.

A summary of the results is shown in Table \ref{tab:objective_rank_table}. As seen in the results, the majority of the systems using ARLM+Diffusion are in the top, showing the effectiveness of the ARLM+Diffusion method in this task. In particular, ablation systems from S6 using supervised fine-tuning, the FiLM-style layer-norm conditioning and a style-aware cross-attention for enhanced fine-grained style modeling are shown to be the most effective out of all the submitted systems, as well as the curation of a high-quality singing dataset.

\section{Conclusions}
The Singing Voice Conversion Challenge 2025 started a new research trend, shifting the focus to not only singer identity conversion, but also singing style conversion. To kickstart this field, we open-sourced baseline systems, tasks, and a dedicated dataset, and also conducted a large-scale evaluation. Results found that although top systems had high scores in singer identity similarity, converting the singing style still has a lot of room for improvement. In particular, we found that creating a system that requires to model both singer identity and singing style at the same time results in lower naturalness scores, and that modeling dynamic information in singing styles such as breathy, glissando, and vibrato needs more work.

A detailed architectural analysis of the top-performing systems in this challenge reveals several crucial best practices for achieving high scores in SSC:
\begin{itemize}
  \item \textbf{Feature disentanglement and fusion.} To highly decouple content and style from singer-related features, it was essential to use disentangled features and model the resulting singing voice in a cascaded fashion. In particular, first, content-style tokens were modeled using a speech LLM from text and prosodic features. Next, using the resulting prosodic features, a diffusion-based model to control the singer timbre was used to generate the acoustic features. These acoustic features were then modeled into a waveform using strong audio vocoders.
  \item \textbf{Fine-grained style conditioning.} Modeling complex dynamic singing styles requires more than simple vector inputs. Modules that dynamically adapt style features deep within the network, such as FiLM-based layer normalization, style-aware Cross-Attention, or residual style adapters, proved extremely effective in capturing both the singer identity and singing style. Adopting LLM post-processing techniques such as DPO was also helpful for further modeling the singing style.
  \item \textbf{Data scale.} Top systems show a robust data curation strategy. For example, Vevo (B2) was pretrained on 100k hours of speech and 7k hours of pretraining data. The submission in S6 further improved this by also focusing on gathering 500 hours of high-quality, noise-filtered singing data.
\end{itemize}

Moreover, we see the importance of continuing running academic challenges such as SVCC. Although commonly used independent metrics such as neural MOS to evaluate naturalness, and non-match metrics such as speaker embeddings to evaluate singer identity are convenient as it does not fully rely on having the ground truth utterances, controlled cases such as when the ground truth data is available in academic challenges allows the use of dependent metrics such as chroma-alignment-related metrics, making benchmarking synthetic singing voices with just objective metrics much easier.

Finally, we see the need to improve both subjective and objective evaluations. We discuss the limitations of both the traditional XAB similarity test and the dynamic preference tests in evaluating the singing styles, showing that there is still a lot of work in improving the subjective evaluations. On the other hand, the SRCC of objective scores and subjective scores only reach a maximum of about 0.8, and thus more work needs to be done before objective tests can become a true replacement for subjective tests. To be specific, finding or creating objective tests that could specifically measure dynamic information could be a promising next direction for the SSC field, due to how these singing styles were more difficult to model by submitted systems.

\noindent\textbf{Acknowledgements} This work was partly supported by JST AIP Acceleration Research JPMJCR25U5 and JSPS KAKENHI (26H02530), Japan.
\section{References}
\printbibliography[heading=none]

\end{document}